\documentstyle[]{article}

\newcommand{\be}{\begin{equation}}
\newcommand{\ee}{\end{equation}}
\newcommand{\bea}{\begin{eqnarray}}
\newcommand{\eea}{\end{eqnarray}}
\newcommand{\CC}{{\cal C}}
\newcommand{\CH}{{\cal H}}
\newcommand{\CR}{{\cal R}}
\newcommand{\CHtil}{\tilde{\cal H}}
\newcommand{\BR}{\bar{R}}
\newcommand{\gbold}{{\bf g}}
\newcommand{\pibold}{{\mbox{\boldmath $\pi$}}}
\newcommand{\betabold}{{\mbox{\boldmath $\beta$}}}
\newcommand{\barnab}{\bar{\nabla}}
\newcommand{\CX}{{\cal X}}
\newcommand{\doh}{\hat{\partial}_0}

\begin{document}
\renewcommand{\thefootnote}{\fnsymbol{footnote}}
\title{CAUSAL PROPAGATION OF CONSTRAINTS AND THE
	CANONICAL FORM OF GENERAL 
	RELATIVITY}
\author{JAMES W. YORK, JR. \\
	{\it Department of Physics and Astronomy,}\\
        {\it University of North Carolina, Chapel Hill, NC 27599-3255} }
\date{\today}
\maketitle
\begin{abstract}
%\abstracts
{Studies of new hyperbolic systems for the Einstein evolution
equations show that the ``slicing density'' $\alpha(t,x)$ can 
be freely specified while the lapse $N = \alpha g^{1/2}$ cannot.
Implementation of this small change in the Arnowitt-Deser-Misner
action principle leads to canonical equations that agree with
the Einstein equations whether or not the constraints are
satisfied.  The constraint functions, independently of their
values, then propagate according to a first order symmetric
hyperbolic system whose characteristic cone is the light cone.
This result follows from the twice-contracted Bianchi identity
and constitutes the central content of the constraint ``algebra''
in the canonical formalism.}
\end{abstract}

In this paper\footnote{Dedicated to Richard Arnowitt} 
I will show one way to follow a path to some small, but
striking
improvements in the Arnowitt-Deser-Misner (ADM) canonical form
of general relativity.~\cite{ADM,Dirac} Arlen Anderson,
collaborator in the primary reference~\cite{AAJWY98} on this
subject, and I followed a different track in the presentation
given elsewhere.~\cite{AAJWY98}

Several aspects of the ADM canonical form of general 
relativity~\cite{ADM,Dirac} have been improved recently.~\cite{AAJWY98}
Slight modifications give rise to possibly significant changes
of perspective. The modifications were suggested by the 
structure of the evolution equations written in new, 
manifestly hyperbolic, forms, not directly relevant to
the present work, in which only physical characteristic
speeds 
appear.~\cite{CBY95,AACBY95,AACBY96,CBY97,AACBY97,AACBYTexas,ACBY97,CBYA98}
One may see some aspects of those
results as flowing to, or from, the action principle.

Here I focus on the consequences of a direct demonstration
that the constraint functions, whether they vanish or not, 
always propagate as dictated by a first order symmetric
hyperbolic system whose only characteristic cone is the full
physical light cone. This system is in fact just a new reading
of the twice-contracted Bianchi identities. This reading 
suggests the form that the canonical momentum evolution 
equations ought to take. This observation, in turn, causes
one to change the treatment of the lapse function N; and 
this change leads to a small alteration in the action 
principle. One's view of general relativity is then changed by
noting that parameter time differentiation of the canonical
variables now holds correctly throughout the entire phase
space, among other things.

The definition of the Einstein tensor 
$G_{\mu \nu} = R_{\mu \nu} - \frac{1}{2} R g_{\mu \nu}$
in terms of the Ricci tensor $R_{\mu \nu}$ leads to the
observation that $2 G^0_0 \equiv R^0_0 - R^k_k$ and to 
the important identity
\be
G_{i j} + g_{i j} G^0_0 \equiv R_{i j} - g_{i j} R^k_k \; .
\label{GIdentity}
\ee
The vanishing of the right hand side does not depend on
any constraints and is equivalent to $R_{i j}=0$. Evidently,
$R_{i j}=0$ is also equivalent to $G_{i j} = -g_{i j} G^0_0$.
Therefore, while $R_{\mu \nu} = 0$ and $G_{\mu \nu} = 0$ are 
equivalent, $R_{i j}=0$ and $G_{i j}=0$ are not equivalent
equations of motion unless the Hamiltonian constraint
\be
\CH \equiv g^{1/2} \CC \equiv 2 g^{1/2} G^0_0
	\equiv g^{1/2} (K_{i j} K^{i j} - H^2 - \BR) = 0
\ee
holds exactly. (Here, $H=K^j_j$ denotes the trace of the 
extrinsic curvature $K_{i j}$ of a spacelike ``time slice''
with metric $g_{i j}$, density 
$g^{1/2} = (\det g_{i j})^{1/2}$, scalar curvature $\BR$,
and spatial covariant derivative $\barnab_i$.)

These observations lead immediately to a transparent form of
the twice-contracted Bianchi identities 
$\nabla_\beta G^\beta_\alpha \equiv 0$. Namely,
\bea
\nabla_\beta G^\beta_0 &\equiv& \nabla_0 G^0_0 +
	\nabla_j G^j_0 \equiv 0 \; , 
\label{Bianchi1}\\
\nabla_\beta G^\beta_j &\equiv& \nabla_0 G^0_j -
	\nabla_j G^0_0 + \nabla_i \left[ R^i_j - 
	\delta^i_j R^k_k \right] \equiv 0 \; .
\label{Bianchi2}
\eea
Recall that the momentum constraints are given by 
\be
\CH_i \equiv g^{1/2} \CC_i \equiv 2 g^{1/2} N R^0_i
	\equiv 2 g^{1/2} N G^0_i = 0 \; .
\ee

Let us write the Bianchi identities (\ref{Bianchi1}),
(\ref{Bianchi2}) in $3+1$ form using the spacetime cobasis
indicated by parentheses in the spacetime metric
\be
ds^2 = -N^2 \left(dt\right)^2 + g_{i j} 
	\left( dx^i + \beta^i dt\right) 
	\left( dx^j + \beta^j dt\right) \; ,
\label{lineelt}
\ee
where $N$ is the lapse function and $\beta^i$ is the shift
vector. The dual vector basis is 
$\partial_0 = \partial_t - \beta^j \partial_j$, with 
$\partial_j \equiv \partial / \partial x^j $ and
$\partial_t = \partial / \partial t$. We define our time 
derivative operator on time-dependent spatial tensors by 
$\doh = \left(\mbox{over-dot}\right)
	= \partial_t - \pounds_\betabold$,
with $\pounds_\betabold$ the spatial Lie derivative along the 
shift vector.  We also use the spatial covariant derivative
$\barnab_j$.

From the cobasis
$\left\{ \theta^0 = dt, \mbox{ } 
	\theta^i = dx^i + \beta^i dt \right\}$
and 
$d\theta^\alpha = -\frac{1}{2} {C^\alpha}_{\beta \gamma}
	\theta^\beta \bigwedge \theta^\gamma$
we find ${C^i}_{0 j} = -{C^i}_{j 0} = \partial_j \beta^i$
and all other structure coefficients vanish. This
gives the connection coefficients 
\bea
{\gamma^i}_{j k} &=& {\Gamma^i}_{j k} 
	= {\bar\Gamma^i}_{j k}         \nonumber\\
{\gamma^i}_{0 j} &=& -N {K^i}_j        \nonumber\\
{\gamma^i}_{j 0} &=& -N {K^i}_j 
	+ \partial_j \beta^i           \nonumber\\
{\gamma^i}_{0 0} &=& N \partial^i N             \\
{\gamma^0}_{i j} &=& -N^{-1} K_{i j}   \nonumber\\
{\gamma^0}_{0 i} &=& {\gamma^0}_{i 0} = 
	\partial_i \log N              \nonumber\\
{\gamma^0}_{0 0} &=& \partial_0 \log N \; , \nonumber
\eea
where $\Gamma$ denotes a Christoffel symbol. 
Our derivative convention is 
$\nabla_\alpha \sigma_\beta = \partial_\alpha \sigma_\beta
	- \sigma_\rho {\gamma^\rho}_{\beta \alpha}$.

Now we can go through a straightforward process of
writing (\ref{Bianchi1}) and (\ref{Bianchi2}) in 
terms of 
$\doh \left( \; \right) 
	= \dot{ \left( \; \right) }$
and $\barnab_j$. For the unweighted constraint
functions $\CC$ and $\CC_i$, the Bianchi 
identities then become
\bea
\dot{\CC} - N \barnab^j \CC_j &\equiv& 
	2 \left( \CC_j \barnab^j N + N H \CC 
	- N K_{i j} \left[ \CR_{i j}\right] \right) \; , 
\label{Constraint1}\\
\dot\CC_j - N \barnab_j \CC &\equiv& 
	2 \left( \CC \barnab_j N + 
	\frac{1}{2} N H \CC_j
	- \barnab^i \left( N \left[ \CR_{i j} 
	\right] \right) \right) \; ,
\label{Constraint2}
\eea
where $\CR_{i j} \equiv R_{i j} - g_{i j} R^k_k$.
Similar results hold for $\CH = g^{1/2} \CC$ and
$\CH_i = g^{1/2} \CC_i$. (Related formulas were 
found by Choquet-Bruhat and Noutchegueme~\cite{CBN}
in studying the evolution of matter sources $\rho^{0 0}$,
$\rho^{0 i}$, where 
$\rho^{\beta \alpha} = T^{\beta \alpha} - 
	\frac{1}{2} g^{\beta \alpha} T^\mu_\mu$.)

The system (\ref{Constraint1}), (\ref{Constraint2}) is
first order symmetric (``symmetrizable'') hyperbolic
with characteristic fields consisting of linear
combinations of $\CC$ and $\CC_i$ propagating along the
light cone.  Hence, if the equations of motion 
$\CR_{i j}=0$ (or $R_{i j}=0$) hold, $\CC$ and
$\CC_i$ remain zero in the physical domain of dependence
associated with the region of the initial data surface
on which they were initially satisfied.  If none of
$\CC$, $\CC_i$, $\CR_{i j}$ is exactly
zero, as is inevitably the case in approximations and
numerical applications, then (\ref{Constraint1}),
(\ref{Constraint2}) show how errors in $\CC$ and
$\CC_i$ are driven along the light cone by terms
linear in these errors and by $R_{i j} - g_{i j} R^k_k$.
(It is straightforward to add a cosmological constant
and a matter source $T^{\alpha \beta}$, such that
$\nabla_\beta T^{\beta \alpha}=0$, to this analysis.)

To draw some lessons for the canonical formalism, let
us first express the $3+1$ evolution equations in 
their standard geometrical form (with zero shift,
Ref. 13; arbitrary lapse and shift, Ref. 14; 
spacetime perspective, Ref. 15):
\be
\dot{g}_{i j} \equiv - 2 N K_{i j} \; ,
\label{gdot}
\ee
\be
\dot{K}_{i j} \equiv N \left( -R_{i j} + \BR_{i j}
	+ H K_{i j} - K_{i k} K^k_j 
	- N^{-1} \barnab_i \partial_j N \right) \; .
\label{Kdot}
\ee

A brief look at (\ref{Kdot}) shows that the term 
$\CR_{i j} = \left( R_{i j} - g_{i j} R^k_k \right)$
that appears in the hyperbolic form (\ref{Constraint1}),
(\ref{Constraint2}) of the Bianchi identities is 
made-to-order to produce an equation of motion for the
ADM canonical momentum
\be
\pi^{i j} = g^{1/2} \left( H g^{i j} - K^{i j} \right)
\ee
that {\it contains no constraints}. Indeed, using
(\ref{gdot}) and (\ref{Kdot}), we obtain the
{\it identity}
\bea
\dot{\pi}^{i j} &\equiv& N g^{1/2} 
	\left( \BR g^{i j} - \BR^{i j} \right)
	- N g^{-1/2} \left( 2 \pi^{i k} \pi^j_k 
	- \pi \pi^{i j} \right) \nonumber\\
	& & + g^{1/2} \left( \barnab^i \barnab^j N
	- g^{i j} \barnab_k \barnab^k N \right)
	+ N g^{1/2} \left[ \CR^{i j} \right] \; .
\label{pidot}
\eea
From (\ref{lineelt}), we have
\be
\dot{g}_{i j} \equiv N g^{-1/2} 
	\left( 2 \pi_{i j} - \pi g_{i j} \right) \; .
\label{gdot2}
\ee
If we now compute the time derivatives of 
$\CC \left(\gbold, \pibold \right)$ and
$\CC_i \left( \gbold, \pibold \right)$
using (\ref{pidot}), (\ref{gdot2}), we obtain, 
of course, the Bianchi identities in the form
(\ref{Constraint1}), (\ref{Constraint2}).  This
merely stresses that the Bianchi identities,
properly construed, are just the hyperbolic 
evolution equations for those phase space functions
whose vanishing yields the ``constraint hypersurface''
in phase space, and in particular we see that 
(\ref{Constraint1}), (\ref{Constraint2}) hold on or
off the constraint hypersurface.

We have now reached a crucial point in the development.
If the canonical equation for $\dot{\pi}^{i j}$ is 
dictated by the vanishing of the spatial part of the
Einstein tensor, $G^{i j} = 0$, as in the ADM 
analysis,~\cite{ADM} then the identities (\ref{GIdentity})
and (\ref{pidot}) show that the constraint term
$\left( 1/2 \right) g^{i j} g^{1/2} \CC \equiv
	\left( 1/2 \right) g^{i j} \CH$
remains in the $\dot{\pi}^{i j}$ equation, restricting
its validity to the subspace of phase space on which
the constraints are satisfied ({\it i.e.}, where the 
constraint functions vanish).  Furthermore, if we
substitute $G^{i j}=0$ back into the Bianchi identities
(\ref{Bianchi1}) and (\ref{Bianchi2}), then the 
hyperbolicity and well-posedness of the constraint
evolution are lost. (The latter has been observed by Frittelli
using different methods.~\cite{Frittelli} She also 
found that $R_{i j}=0$ gives well-posed evolution.)

However, though the ADM derivation of the $\dot{\pi}^{i j}$
equation, found by varying $g_{i j}$ in their canonical
action ($16 \pi G = c = 1$)
\be
S \left[ \gbold, \pibold ; N, \betabold, \right) 
	= \int d^4x \left( \pi^{i j} \dot{g}_{i j}
	- N \CH \right) \; ,
\ee
with $N(t,x)$ and $\beta^i(t,x)$ as undetermined 
multipliers, is of course perfectly correct, another point
of view is possible.  (We are ignoring boundary terms,
which are not of interest here, and we note that the
momentum constraint term $-\beta^i \CH_i$ 
($\CH_i = g^{1/2} \CC_i$) is contained in
$\pi^{i j} \dot{g}_{i j}$ 
($\dot{\left( \; \right)} \equiv \doh $)
upon integration by parts.) The other point of view 
arises in our work on hyperbolic systems with only
{\it physical} 
characteristics.~\cite{CBY95,AACBY95,AACBY96,CBY97,AACBY97,AACBYTexas,ACBY97,CBYA98}
There, one is {\it never} allowed to specify $N(t,x)$
freely! Rather one must use in essence Choquet-Bruhat's
``algebraic gauge,'' which asserts that the weight-minus-one
lapse (the ``slicing density'') 
$\alpha = N g^{-1/2} = \alpha(t,x) > 0$ is {\it freely} 
specifiable while $N$ is not.~\cite{CBR,CBY95}
(The slicing density $\alpha$ has been used prominently
in the action by Teitelboim~\cite{Teitel82} and 
Ashtekar~\cite{Ashtekar88,Ashtekar87} for other 
purposes.)

Indeed, if one computes 
$\doh \log \alpha = f(t,x)$
from a given $\alpha(t,x)$, then
\be
g^{1/2} \doh \alpha 
	= \doh N + N^2 H = N f \; ,
\label{Harmonic}
\ee
which is the equation of harmonic time slicing with
$f(t,x)=\doh \log \alpha$ acting as a 
``gauge source.''~\cite{Friedrich}  In this sense
the undetermined multiplier $\alpha(t,x)$, like
$\beta^i(t,x)$, cannot affect the issue of
hyperbolicity.~\cite{AAJWY98,CBY95,AACBY95,AACBY96,CBY97,AACBY97,AACBYTexas,ACBY97,CBYA98}
The ``harmonic time slicing gauge'' is not a gauge
at all; it is simply the equation of motion of the 
lapse function (with a ``gauge source'') once one 
recognizes $\alpha$ as the true undetermined multiplier.

Combining (\ref{Harmonic}) and the equation for 
$\dot{H}$ obtained from (\ref{Kdot}), one obtains 
a quasi-linear wave equation (hyperbolic with 
characteristic speed $c=1$!) for $N$.  Therefore,
{\it every} foliation of a globally hyperbolic 
spacetime by regular time slices is given by some
initial time slice, and the solution $N>0$ of a wave
equation, for some $\alpha(t,x) > 0$.  {\it No} slices
are ``lost'' by changing from $N$ to $\alpha$, so the
latter is as ``good'' as the original $N$. (Degenerate
cases can be handled as well with $g^{1/2}$ and 
$\alpha$ as with $N$.)  We conclude that $N$ is a
{\it dynamical variable} (as did 
Ashtekar~\cite{Ashtekar88,Ashtekar87} for other reasons) 
which determines
the proper time $N \delta t$ between slices $t=t^\prime$
and $t=t^\prime+\delta t$. It is determined from 
$\alpha(t,x)$ and the solution $g^{1/2}$ of the
constraint equations.~\cite{York79,York73,OMYork,CBY80}
From this perspective, the Hamiltonian constraint plays
its familiar role as an initial-value constraint that
determines $g^{1/2}$ given the other free data.~\cite{York79}
The important point is that $\CH=0$ does not
determine the time but does fix the rate of proper time
$\tau$ with respect to parameter time $t$:
$d\tau/dt = g^{1/2} \alpha$ along the normal $\partial_0$.

Motivated by the above findings and those with respect
to the $\dot{\pi}^{i j}$ equation and the well-posedness
of the Bianchi equations as a causal hyperbolic system,
we alter the undetermined multiplier $N$ in the ADM
action principle to $\alpha$. Thus the Hamiltonian
function becomes
\be
\CHtil = g^{1/2} \CH = 
	\pi_{i j} \pi^{i j} - \frac{1}{2} \pi^2 - g \BR \; ,
\ee
where $\CHtil$ is a rational function of the metric
of scalar weight plus two.  The 
ADM-plus-Teitelboim-plus-Ashtekar action becomes
\be
S\left[\gbold, \pibold; \alpha, \betabold \right) =
	\int d^4 x \left( \pi^{i j} \dot{g}_{i j}
	- \alpha \CHtil \right) \; .
\label{action}
\ee
The modified action {\it principle} for the canonical
equations that we propose is to vary $\pi^{i j}$ and
$g_{i j}$, with $\alpha(t,x)$ and $\beta^i(t,x)$ as
undetermined multipliers. From
\bea
\delta \CHtil &=& \left( 2 \pi_{i j} - g_{i j} \pi \right)
	\delta \pi^{i j} 
	+ \left( 2 \pi^{i k} \pi^j_k - \pi \pi^{i j} 
	+ g \BR^{i j} - g g^{i j} \BR \right) \delta g_{i j} 
	\nonumber\\
	& & - g \left( \barnab^i \barnab^j \delta g_{i j}
	- g^{i j} \barnab_k \barnab^k \delta g_{i j} \right)
\eea
we obtain the canonical equations
\bea
\dot{g}_{i j} = \alpha \frac{\delta \CHtil}{\delta \pi^{i j}}
	&=& \alpha \left( 2 \pi_{i j} - \pi g_{i j}\right)
	\equiv -2 N K_{i j} \; , \\
\label{Canonicalgdot}
\dot{\pi}^{i j} = -\alpha \frac{\delta \CHtil}{\delta g_{i j}}
	&=& - \alpha g \left( \BR^{i j} - \BR g^{i j} \right )
	- \alpha  \left( 2 \pi^{i k} \pi^j_k - \pi \pi^{i j} \right)
	\nonumber\\
	& & + g \left( \barnab^i \barnab^j \alpha - 
	g^{i j} \barnab_k \barnab^k \alpha \right) \; .
\label{Canonicalpidot}
\eea
Equation (\ref{Canonicalpidot}) for $\dot{\pi}^{i j}$
is the identity (\ref{pidot}) with $\CR^{i j}=0$,
that is, $R^{i j}=0$. Thus, (\ref{Canonicalpidot}) is a 
``strong'' equation unlike its ADM counterpart which requires
in addition the strict validity of a constraint: $\CH=0$.

In the present formulation, the canonical equations of 
motion hold everywhere on phase space with any 
parameter time $t$, a necessary condition for the issue
of ``constraint evolution'' to be discussed properly
at all in the Hamiltonian framework.  Further, there
seems to be no need to vary $\alpha$ and
$\beta^i$ in a spacetime action volume integral
to enforce the constraints for {\it all} time, because the constraints
are dynamically determined to hold in the appropriate
physical domain of dependence if they hold initially, 
as shown below in (\ref{newConstraint1}), (\ref{newConstraint2}).  
Perhaps $\alpha$ and $\beta^i$ should be varied on an
initial slice only.

If we define the integrated or ``smeared'' Hamiltonian
constraint as
\be
\CHtil_\alpha = \int d^3 x^\prime \alpha(t,x^\prime) \CHtil \;,
\ee
the equation of motion for a general functional
$F[\gbold,\pibold;t,x)$ anywhere on the phase space
is 
\be
\dot{F}\left[ \gbold, \pibold; t, x \right) = 
	- \left\{\CHtil_\alpha, F\right\} 
	+ \tilde{\partial}_0 F \; ,
\label{FunctionalEOM}
\ee
where $\dot{\left( \; \right)}$ denotes our total time
derivative and $\tilde{\partial}_0$ is a ``partial'' 
derivative of the form $\partial_t - \pounds_\betabold$
acting only on explicit spacetime dependence.  The 
Poisson bracket is
\be
\left\{F,G\right\} = \int d^3 x 
	\left( \frac{\delta F}{\delta g_{i j}(t,x)}
	\frac{\delta G}{\delta \pi^{i j}(t,x)}
	- 
	\frac{\delta G}{\delta g_{i j}(t,x)}
	\frac{\delta F}{\delta \pi^{i j}(t,x)} \right) \; .
\ee
It is clear that the $\left( \dot{\gbold}, \dot{\pibold} \right)$
equations come from (\ref{FunctionalEOM}) applied to the
canonical variables. The harmonic time slicing equation
(\ref{Harmonic}) results from application of (\ref{FunctionalEOM})
to $N$, and the wave equation for $N$ comes from a repeated 
application of (\ref{FunctionalEOM}) to (\ref{Harmonic}).

%\clearpage
Time evolution is generated by the Hamiltonian vector field
\bea
\CX_{\CHtil_\alpha} &=& \int d^3 x \left\{ 
	\alpha ( 2 \pi_{i j} - \pi g_{i j} )
	\frac{\delta}{\delta g_{i j}} 
	- [ \alpha g ( \BR^{i j} 
	- \BR g^{i j}) \right.
\nonumber\\
	& & + \left. \alpha ( 2 \pi^{i k} \pi^j_k 
	- \pi \pi^{i j} )
	- g ( \barnab^i \barnab^j \alpha 
	- g^{i j} \barnab_k \barnab^k \alpha ) ] 
	\frac{\delta}{\delta \pi^{i j}} \right\} \; .
\label{HamVecField}
\eea
Because is does not contain any explicit constraint
dependence, (\ref{HamVecField}) is a valid time evolution
operator on the entire phase space.

The product rule applicable to $\doh$ and the Poisson
bracket shows that the evolution equations 
(\ref{Constraint1}), (\ref{Constraint2}) for the
constraints can be written as
\bea
\doh\CHtil &=& - \left\{ \CHtil_\alpha, \CHtil \right\}
	= \alpha g g^{i j} \partial_i \CH_j 
	+ 2 g g^{i j} \CH_i \barnab_j \alpha \; , 
\label{newConstraint1}\\
\doh\CH_j &=& - \left\{ \CHtil_\alpha, \CH_j \right\}
	= \alpha \partial_j \CHtil 
	+ 2 \CHtil \partial_j \alpha \; ,
\label{newConstraint2}
\eea
where $\barnab_j \alpha = \partial_j \alpha 
	+ \alpha g^{-1/2} \partial_i g^{1/2}$.
Therefore, the Poisson brackets of the smeared with the
unsmeared constraints are well-posed evolution 
equations for the constraints.  These equations
are, of course, just the twice-contracted Bianchi
identities when $\CR_{i j}=0$ or $R_{i j}=0$.

The consistency of time evolution with respect to
different choices of $\alpha$ is checked by using
the Jacobi identity,
\be
\left\{ \CHtil_{\alpha_1}, \left\{ \CHtil_{\alpha_2}, 
	F \right\} \right\} -
\left\{ \CHtil_{\alpha_2}, \left\{ \CHtil_{\alpha_1}, 
	F \right\} \right\} =
\left\{ \left\{ \CHtil_{\alpha_1}, \CHtil_{\alpha_2}  
	\right\}, F \right\} \; , 
\label{Jacobi1}
\ee
and
\be
\left\{ \CHtil_{\alpha_1}, \CHtil_{\alpha_2} \right\}
	= - \int d^3 x g g^{i j} 
	\left( \alpha_1 \partial_i \alpha_2 
	- \alpha_2 \partial_i \alpha_1 \right) \CH_j \; .
\label{Jacobi2}
\ee
The metric dependence of (\ref{Jacobi2}) shows that
the difference between evolution with $\alpha_2$
followed by $\alpha_1$, and {\it vice versa}, is
a spatial diffeomorphism when $\CH_i = 0$ or when
$\delta F/\delta \pi^{i j}=0$.

The results of the previous paragraphs shed new light
on the Dirac ``algebra'' of constraints (cf. Ref~25).
It is well known that the
Dirac algebra is not the spacetime diffeomorphism
algebra.  This can be seen from the fact that
while the action (\ref{action}) is invariant under
transformations generated by the constraint 
functions even when they do not vanish,~\cite{Teitel77}
the equations of motion that follow from this action
when the constraint functions do not vanish
are $R_{i j}=0$.  These equations are preserved by
spatial diffeomorphisms and time translations along
their flow in phase space, but a general spacetime
diffeomorphism applied to $R_{i j}=0$ mixes in the 
constraints.  A comparison of (\ref{Constraint1}),
(\ref{Constraint2}) with (\ref{newConstraint1}),
(\ref{newConstraint2}) shows this effect: the
Bianchi identities are spacetime diffeomorphism
covariant while (\ref{newConstraint1}), 
(\ref{newConstraint2}) are not. The equations
(\ref{Constraint1}), (\ref{Constraint2}) and
(\ref{newConstraint1}), (\ref{newConstraint2})
differ precisely by terms proportional to 
$\CR_{i j}$ and $\barnab^i \CR_{i j}$.

A second important view of the Dirac algebra
results from the direct and beautiful dynamical
meaning of its once-smeared form.  Equations
(\ref{newConstraint1}) and (\ref{newConstraint2})
express consistency of the constraints as a well
posed initial-value problem. If the constraint
functions vanish initially, they continue to do
so by evolution into the domain of dependence 
corresponding to the region of the initial time slice
on which they initially vanished.  This mechanism
follows from the dual role of $\CHtil$ as a 
constraint and as part of the generator of
time translations of functionals of the canonical
variables anywhere on the phase space.

Let us take note that the Hamiltonian constraint
{\it per se} does not express the dynamics of 
the theory; the equation of dynamics is 
(\ref{FunctionalEOM}).  In its ``altered'' 
role, the Hamiltonian constraint function simply
vanishes as an initial value condition, from which 
$g^{1/2}$ is determined as in the intial value
problem.~\cite{York79} Then $N$ can be constructed
from $\alpha$. The Hamiltonian constraint, once
solved, remains so according to the rigorous result
embodied in (\ref{newConstraint1}), 
(\ref{newConstraint2}).

The application of these ideas to canonical quantum
gravity will appear elsewhere.~\cite{AAJWYnew}

\section*{Acknowledgments}
The author thanks Arlen Anderson, co-author of
the primary reference,~\cite{AAJWY98} for many
helpful ideas and insights.  This work was 
supported by National Science Foundation 
grants PHY~94-13207 and 
PHY~93-18152/ASC~93-18152 (ARPA supplemented).

\end{document}